\documentclass{IEEEtran}
\usepackage[utf8]{inputenc}
\usepackage{hyperref}
\usepackage{amsmath,amssymb,amsfonts}
\usepackage{graphicx}
\usepackage{upgreek}
\usepackage{xcolor}

\makeatletter
\newlength{\myfigurewidth}
\if@twocolumn
\else
\fi
\makeatother

\title{Automatic calibration of gamma-ray detectors deployed in uncontrolled environments}

\author{Marco Salathe, Nicolas Abgrall, Mark S.\ Bandstra, Tenzing H.\ Y.\ Joshi, Brian J.\ Quiter and Reynold J.\ Cooper
\thanks{
This work was performed under the auspices of the U.S. Department of Energy by Lawrence Berkeley National Laboratory (LBNL) under Contract DE-AC02-05CH11231. The project was funded by the U.S. Department of Energy, National Nuclear Security Administration, Office of Defense Nuclear Nonproliferation Research and Development.
}
\thanks{
All authors are with the Applied Nuclear Physics Program at Lawrence Berkeley National Laboratory, Berkeley, CA 94720 USA. Email: msalathe@lbl.gov.}
}

\begin{document}
\bstctlcite{IEEEexample:BSTcontrol}
\maketitle

\begin{abstract}
    Radiation detectors deployed as part of a large urban network or for homeland security monitoring must maintain reliable energy calibration even when subjected to substantial variations in temperature and ambient background radiation.
    Traditional calibration methods often rely on power-intensive temperature stabilization or peak-locking algorithms that are susceptible to environmental changes.
    This publication presents a novel software-based calibration method that eliminates the need for active temperature control by utilizing full-spectrum analysis.
    The method continuously updates the calibration parameters by fitting the spectral data with a series of background radiation contributions (K, U, Th series, radon progeny and cosmics) combined with a Monte-Carlo-based physical detector model that incorporates light yield non-proportionality and photomultiplier tube saturation.
    Performance was validated using simulated data, measurements in an environmental chamber across a wide temperature range ($-25^\circ$C to $+50^\circ$C), and data from a multi-day outdoor field deployment.
    Results demonstrate that the method successfully maintains stable energy calibration despite significant ambient temperature variations and precipitation events.
    The technique effectively decouples instrumental drift from spectral changes caused by environmental background fluctuations.
    This approach provides a robust, automated, and low-power alternative to conventional calibration techniques, enabling the practical deployment of large-scale, unattended networked detector systems.
\end{abstract}

\section{Introduction}

\IEEEPARstart{T}{he} deployment of distributed sensor networks in natural and urban environments represents a significant paradigm shift for nuclear non-proliferation and homeland security.
Traditional radiation monitoring has historically relied on discrete, localized ``choke point'' systems at ports of entry, high-traffic areas, or borders \cite{Kouzes2020}.
In contrast, networked architectures provide a cohesive framework for continuous, wide-area monitoring that covers entire cities or other large areas of interest.
Beyond detecting nuclear materials in high-traffic areas, these systems can be adapted to a wide array of mission spaces—from monitoring perimeters around nuclear power plants to characterizing radiological migration in contaminated areas during remediation efforts \cite{international2005iaea}.
By integrating low-resolution gamma-ray detectors with contextual sensors like cameras and lidar, these networks reduce the operational burden of manual oversight and enable the direct correlation of radiological signatures with specific objects or dynamic changes in the scene.
Ultimately, such a network is not limited to observing objects at discrete locations, but instead enables the persistent tracking of radiological materials across the entire monitored environment.
The Platforms and Algorithms for Networked Detection and Analysis (PANDA) project \cite{Cooper2023} is one such effort aiming to develop these new technologies, with a focus on a city-scale deployment of networked sensors.
The presented work is part of this project.

A critical challenge for any radiation detection system operated outside of a radio-pure ``laboratory'' environment is distinguishing an illicit source from the often complex and variable backgrounds.
The main contributions to background radiation are naturally occurring radioactive materials (NORMs) in dirt, rock, and building materials (e.g., K-40, the U-238 and Th-232 decay series), cosmic rays, and airborne radioactive dust.
These backgrounds can obscure the presence of human-made radioactive material and NORM that has been concentrated to unsafe levels.

To overcome this limitation, many advanced systems rely on spectroscopy --- using a detector's sensitivity to energy deposited by gamma-ray interactions --- to search for distinct emission signatures from known sources.
This specificity increases the device's sensitivity by enabling algorithms that leverage detailed spectral information in more sophisticated ways \cite{10247022, PFUND2016174, 8673769, Mitchell2013GADRAS}.

However, these algorithms are perturbed by detector instabilities.
Environmental factors can cause the detector's response to drift, making gamma-ray peaks appear at the wrong energies. 
Stabilizing the location of these peaks is essential for any unattended, deployed spectrometer.
While these instabilities are similar in nature for different types of detectors, the presented work focuses on scintillators, particularly Thallium-loaded sodium iodide, or NaI(Tl) detectors, for which changes in the ambient temperature are the dominant effect \cite{ianakiev2006temp}.

In the past, two approaches have been deployed to stabilize the energy response.

The first method is active temperature control, which uses heating and cooling elements to stabilize the detector at a specific temperature where the calibration is already established.
This approach is energy-intensive, requires complex engineering to handle wide temperature swings (e.g., -20°C to 40°C), and can even worsen energy resolution in scintillators like NaI(Tl) by reducing light yield \cite{ianakiev2006temp}.

The second strategy is automated gain tracking, where software adjusts the detector's operational parameters (like bias voltage) to keep spectral features ``locked'' in place.
Typically, this process monitors one or two prominent NORM gamma-ray lines.
However, this approach constrains the correction to a linear model, which fails to stabilize the full spectral range and leaves substantial shifts at lower energies.
It can also fail if the background lines are obscured by contamination or if a peak drifts too far out of its monitoring window.
A common variant involves adding a dedicated radioactive check source to provide a stable signal for calibration, but this can obfuscate the energy region of interest.

Both existing methods add significant engineering complexity or operational burden, making them ill-suited for robust, low-maintenance, distributed networks.

Here, a novel method is presented that leverages advanced optimization techniques in software to automatically calibrate spectra with little intervention by an operator and no requirement for temperature control.
The method is inspired by traditional peak-tracking algorithms but makes use of features across the entire spectrum rather than isolated peaks.
This approach makes the calibration far more stable with respect to drifts and initial conditions, and provides a path toward robust, unattended, and scalable networked detection systems.

The new calibration method is described in Section~\ref{sec:methods}, and then evaluated in Section~\ref{sec:results} using simulated data, environmental chamber measurements, and a multi-day outdoor field deployment.
Some observations from these tests are discussed in Section~\ref{sec:discussion} before concluding the paper with Section~\ref{sec:conclusion}.

\section{Methods}
\label{sec:methods}

The present discussion focuses on NaI(Tl) detectors of a relatively large volume read out by a photomultiplier tube (PMT).
All data presented in this paper were gathered with 2$\times$4$\times$16\,inches detectors produced by Saint-Gobain \cite{nai_manual} and read out with an ORTEC/AMETEK digiBASE 14-Pin PMT base \cite{digitbase_manual}.
While the specific assumptions made here are tailored to NaI(Tl) PMT detectors, the general methodology can be adapted to other detector technologies.

\subsection{Detector calibration model}

\begin{figure}
  \centering\includegraphics[width=1.0\myfigurewidth]{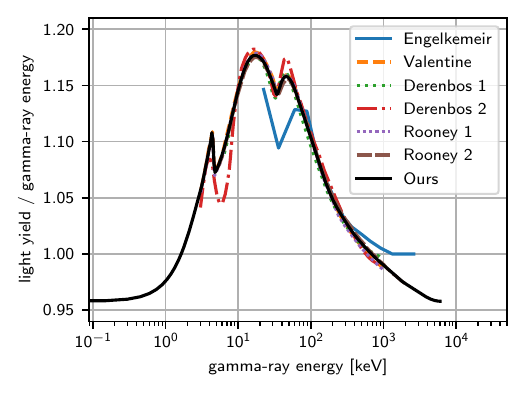}
  \caption{Normalized light yield non-proportionality curves for NaI(Tl) as a function of deposited energy (logarithmic scale). Curves derived from or attributed to Engelkemeir \cite{Engelkemeir1956}, Valentine et al.\ \cite{Valentine1998},  Rooney et al.\ \cite{Rooney1997}, and Dorenbos et al.\ \cite{Dorenbos1995} are compared. The ``Ours'' curve represents a spline fit through various models. Note the characteristic structures near the iodine K and L edges.}
  \label{fig:lightyield}
\end{figure}

Gamma rays deposit energy in a detector by photoelectric absorption, Compton scattering or pair production.  This energy deposition is converted into low-energy scintillation photons by the scintillator material.
The response of the scintillator as a function of the deposited energy is linear to first order but has complex behavior at higher orders.
This complexity is dominated by light yield non-proportionality, or the variation in light output per unit of deposited energy.
Although light yield changes considerably with temperature, measurements of non-proportionality across wide temperature ranges are scarce. For simplicity, we assume the shape of the non-proportionality curve remains constant with temperature.

The light yield model $f_\mathrm{LY}(E)$ used in this publication is derived from early measurements, often performed under well-controlled conditions.
Engelkemeir first highlighted the characteristic increase in relative light yield at lower energies \cite{Engelkemeir1956}.
Subsequent studies have refined our understanding: Valentine and Rooney\ \cite{Valentine1998, Rooney1997} clarified the terminology between linearity and proportionality and developed methods to calculate the photon response from the underlying electron response, showing its dependence on crystal geometry.
Dorenbos et al. \cite{Dorenbos1995} linked non-proportionality to intrinsic energy resolution limits.
Moszynski et al. \cite{Moszynski2002} further investigated the components of intrinsic energy resolution, isolating the potentially contribution from delta-rays by comparing measurements with non-proportionality models.
Importantly, Hull et al. \cite{Hull2009} demonstrated significant sample-to-sample variations in the electron response, particularly below 20 keV.
Some of the light-yield curves proposed in these publications are shown in Figure~\ref{fig:lightyield}.
Although Hull et al. present numerous curves, they are not detailed individually here.
The ``Ours'' curve, which is the model used throughout this work, was generated by aggregating digitized data from several key studies \cite{Valentine1998, Rooney1997,Dorenbos1995}.
A cubic B-spline least-squares fit was applied to this combined dataset.
The resulting spline was then extrapolated to cover the required range using a cubic Hermite spline.

Thus, using this model  for a given deposited energy, $E$, the total light output is given by $L(E) = E \cdot f_\mathrm{LY}(E)$.

In most scintillator detector configurations, the emitted photons are collected at one end of the scintillator connected to a photomultiplier tube (PMT), possibly after several internal reflections against a mirrored surface covering the scintillator material. In the PMT, electrons knocked off a photo-cathode by the emitted photons are amplified in several stages to produce an electronic signal on output. 

While a PMT response is specified to be linear over several orders of magnitude, some non-linearity occurs by design.
For a low intensity pulse the PMT will produce an output which is linear to the input, in proportion to the PMT gain.
A very intense pulse will produce a constant output at the saturation voltage.
The response of the PMT does not suddenly switch from linear to saturated, but instead shifts gradually between the two modes.
The response model for the PMT is best described as a gain and saturation model:

\begin{equation}
    f_\mathrm{PMT}(x) = \frac{g_\mathrm{PMT}\cdot x}{1+k_{\mathrm{PMT}}\cdot x}
\end{equation}

The equation expresses the PMT light output as a function of the total light output of the scintillator, $x$, linking them with a gain factor, $g_\mathrm{PMT}$ and a saturation factor, $k_\mathrm{PMT}$.
The gain of the PMT strongly depends on the applied high-voltage bias.
The saturation, on the other hand, does not increase significantly with the applied voltage.

The next stage following the PMT is an analog amplification and filtering system.
The filter is responsible for converting the rise from photon arrival into a pulse with a height proportional to the deposited energy.
This pulse height is then held by a peak hold circuit for conversion with an analog to digital converter (ADC).

Each of these elements has an electronic gain which we can lump together as the data acquisition system (DAQ) gain.
The shaper removes the baseline of the system to define a zero level from which the pulse height is measured.
However, both the peak hold circuit and the ADC may have an offset.

The effect of the data acquisition system can thus be summarized with the equation:

\begin{equation}
    f_\mathrm{DAQ}(x) = g_\mathrm{DAQ} x + o_\mathrm{DAQ}
\end{equation}

where the DAQ output, fDAQ, is obtained by multiplying the input signal with a gain factor $g_\mathrm{DAQ}$ and adding a constant offset $o_\mathrm{DAQ}$.

The complete system response that  provides the pulse height, C, as a function of the deposited energy is then given by the composition of these elements:

\begin{equation}
    C = f_\mathrm{DAQ}(f_\mathrm{PMT}(L(E))),
\end{equation}
which can be simplified to a model with an overall gain $g=g_\mathrm{PMT}\cdot g_\mathrm{DAQ}$, saturation $k=k_\mathrm{PMT}$ and offset $o=o_\mathrm{DAQ}$:

\begin{equation}
    C = \frac{g \cdot L(E)}{1 + k\cdot L(E)} + o
\end{equation}

The relation can be inverted to express the energy $E$ as a function of the observed pulse height $C$:

\begin{equation}
    E = L^{-1}\left(\frac{(C - o)}{g - k \cdot (C - o)}\right)
    \label{eq:calibration}
\end{equation}

In Equation~\ref{eq:calibration}, the term $L^{-1}$ represents the inverse of the total light function $L(E) = E \cdot f_\mathrm{LY}(E)$ as previously defined.
As this function cannot be inverted analytically, its inversion is implemented numerically.
A lookup table is pre-computed from the spline, mapping total light output $L$ back to the corresponding deposited energy $E$.
The function $L^{-1}$ is then executed as a fast linear interpolation on this table.

The model thus can be summarized as a three parameter model that converts pulse heights to energies.

\subsection{Energy resolution model}

The exact energy of a gamma ray is smeared by various effects of the signal conversion, amplification and digitization chain.
There are various models and other equations used for modeling the energy resolution as a function of energy \cite{CASANOVAS201278}.

The simplest model is the square root of energy relation.
This relation derives from the statistical nature of the number of particles produced in the conversion and amplification steps.
The smallest number of particles observed in any of the processes is usually considered decisive with $\Delta N / N \approx 1 / \sqrt{N}$ limiting the energy resolution.
The number of particles $N$ is proportional to the deposited energy, thus the effective resolution, usually expressed as a full-width at half-maximum (FWHM), is proportional to the square root of the energy.
A detailed understanding of the energy resolution model is not the primary goal of this work and it has been observed that the details of the implemented model do not affect the quality of the resulting calibration considerably.
A variation of the simple square root relationship between energy and resolution is used.
The model is parameterized by the fractional FWHM resolution at 662\,keV, which is denoted as $\alpha$.
This energy corresponds to a prominent gamma-ray transition observed during the decay of Cs-137, which is often quoted in the literature.
The FWHM at any energy $E$ is then given by:
\begin{equation}
    \mathrm{FWHM}(E) = \alpha \sqrt{E / 662\,\mathrm{keV}}
    \label{eq:resolution}
\end{equation}
This parameter $\alpha$ (e.g., 0.08 for 8\% energy resolution) is the value used in the optimization bounds in Table~\ref{tab:bounds}.

\subsection{Modeling of naturally occurring radioactive materials}

Detectors deployed in urban environments experience gamma-ray backgrounds that primarily originate from trace radioactive materials present within the surrounding soil and construction materials, those that are emitted by radon progeny dissolved in rainwater and suspended in the atmosphere, and those that result from cosmic-ray interactions in the atmosphere.
The primary sources of terrestrial gamma rays are the decays of K-40 and isotopes in the U-238 and Th-232 decay series (which will be referred to as ``KUT''), while the main gamma-ray emitting radon progeny are Pb-214 and Bi-214.
A more complete discussion about the behavior of radon and its progeny during rain can be found in \cite{Bandstra2026}.
Cosmic ray interactions lead to a power-law continuum~\cite{sandness_accurate_2009} and a 511\,keV positron annihilation peak~\cite{bandstra_modeling_2020}.

Some combination of terrestrial, atmospheric, rain-based, and cosmic gamma-ray signals is always present in a detector deployed in an urban landscape.
Our method requires a full-spectrum model of the background, and therefore we need quality templates for each background source type across the entire spectrum.

Full-spectrum modeling of these different background source types is challenging.
As the gamma rays travel from their point of emission, any interactions with the environment cause them to lose energy and contribute to the continuum portion of the spectrum.
In principle, even photons from several hundred meters away can contribute non-negligibly to the spectrum measured by a detector, and the more distant photons, because of their likelihood of undergoing multiple scatters, tend to contribute to the lowest energies ($\lesssim$200\,keV).
To properly model the entire gamma-ray spectrum, the soil and atmosphere in Monte Carlo simulations must extend out to sizes of at least 300\,m away from the detector in the horizontal and vertical directions, and with a soil thickness of 1.5\,m~\cite{sandness_accurate_2009}, although others have performed simulations out to at least 1000\,m away~\cite{osti_1805000,bandstra_modeling_2020}.

Because of the extent of the simulation volume relative to the detector size, it is computationally impractical to perform a ``brute-force'' simulation for the various background types.
Both Ref.~\cite{bandstra_modeling_2020} and \cite{sandness_accurate_2009} discuss different methods to solve this ``scale'' problem.
Our method for simulating full-spectrum gamma-ray backgrounds uses a two-stage method described in detail in Ref.~\cite{Bandstra2026}.
The method relies on two insights.
The first is that if we had a planar geometry of soil and atmosphere that extended infinitely in all directions, and the various background sources were emitted uniformly within their relevant media (soil for KUT, the ground surface for rain-related isotopes, the atmosphere for cosmics), then the radiation field at a fixed height above ground would be translationally invariant for every type of background source.
The second insight is that we are interested in the background observed in a detector that is only tens of centimeters long located a few meters above the ground.
Because of its relatively small scale, its perturbation to the radiation field around it can be assumed to be small (except for self-attenuation, which will be accounted for in the simulations).
Furthermore, due to the detector's small vertical extent and the gradual change in the radiation field with height, we assume the radiation field across its volume is constant.

As in Ref.~\cite{Bandstra2026}, the first-stage simulation is used to estimate the radiation field at a flat tally layer.
The tally layer does nothing except register the energies, directions, and particle types of any particles that cross it, traveling both upwards and downwards.
The tallied particle data allow us to estimate the mean radiation field at the height of the detector.
This field is then used as the source of photons for the second stage, which consists of a 25\,cm-radius sphere of air surrounding the detector.
The analysis of the second stage simply creates a histogram of the event energies detected by the detector.
While not relevant here, by accounting for any dimensional factors in the process, it is possible to relate source activity concentrations to the measured spectral count rate densities.

There are some details in the simulation used here that are different from Ref.~\cite{Bandstra2026}.
We used an atmosphere extending 2\,km above the ground surface, and the entire geometry was a cylinder with a radius of 1000\,km.
The radius of 1000\,km was arbitrary but chosen so that the edge effects could be ignored for simplicity.
The tally layer was placed at a height of 2\,meters above the ground, which is the assumed detector height, although in this stage of the simulation no detector is present.

The KUT source photons were generated uniformly from within the soil volume, rain-related source photons were generated uniformly from a thin 0.1-mm layer just above the ground surface, and the 511\,keV photons due to cosmic-ray interactions in the atmosphere were generated uniformly within the entire atmosphere volume.
Ref.~\cite{Bandstra2026} simulates the radon progeny emerging from a thin layer inside the top of the soil volume to approximate surface roughness effects, but that was not done here.

Finally, the process results in five simulated spectral templates: three for the terrestrial background (K-40, U-238 series, Th-232 series), one for rain-related radon progeny (Pb-214/Bi-214 assumed to be in secular equilibrium, although that is often not the case~\cite{Bandstra2026}), and one for the 511\,keV cosmic annihilation line.
Unlike Ref.~\cite{Bandstra2026}, the fallout isotopes Cs-137 and Be-7 were not included here.

An additional analytical component is included to model the cosmic-ray continuum, which is not captured by the 511\,keV simulation. This component is a power law of the form $f(E) \propto E^{-\gamma}$, where the power-law index $\gamma$ is treated as a free parameter in the fitting routine~\cite{sandness_accurate_2009,bandstra_modeling_2020}.
To prevent the fit from being dominated by high flux at low energies (which goes towards infinity), this template is normalized such that the sum of its bins above the user-specified low-energy threshold (e.g., 380\,keV) is equal to one.
All other templates are normalized to one.
That procedure ensures that the fit weights match roughly the number of counts per time window in each component.

This addition brings the total to six spectral components (five simulated, one analytical), all of which are calculated for the energy range 0--4\,MeV with 1\,keV bin widths.

An example of the resulting spectral components is shown in panel (A) of Figure~\ref{fig:steps}.

\subsection{Data modeling}

\begin{figure*}
  \centering\includegraphics[width=1.0\textwidth]{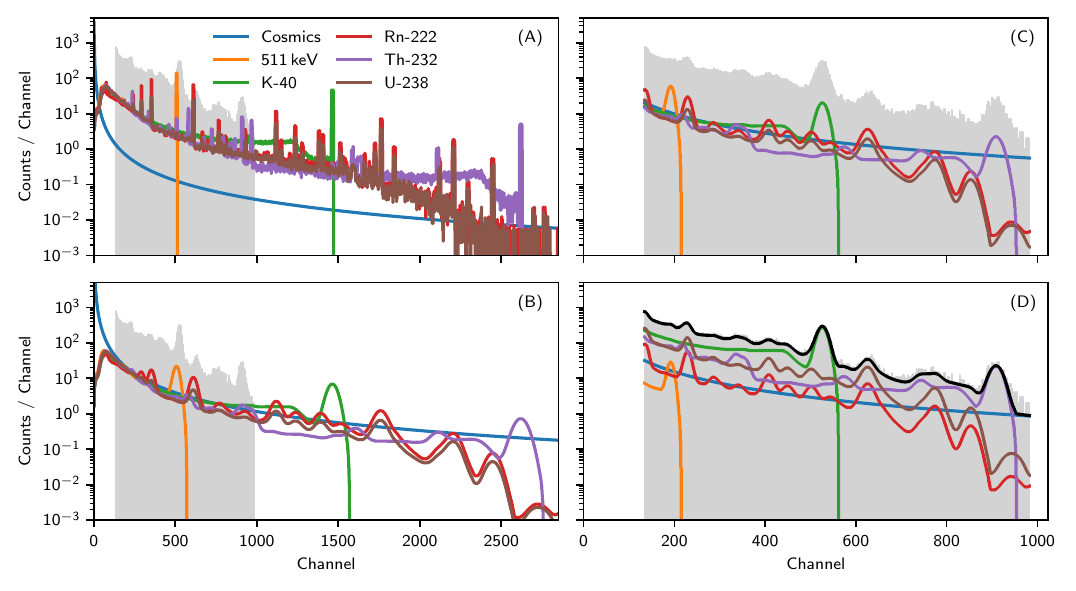}
  \caption{Steps involved in calibrating a single spectrum: (A) The histogram (grey) and the different background components (colors) in their unaltered form. (B) The components are ``smeared'' to match the energy resolution of the detector. (C) The components are squeezed to match the spectrum along X. (D) The amplitude of the components is adjusted (in Y) to sum up to the spectrum.}
  \label{fig:steps}
\end{figure*}

The 6 simulated spectral components are distorted in 3 distinct steps to approximate the data.
First, the simulated templates are broadened to match the detector's energy-dependent resolution (Equation~\ref{eq:resolution}) by using a variable-width, windowed Gaussian filter.

The peak broadening procedure was not applied to the component carrying the power law distribution from cosmic rays.
The resulting templates are displayed in the panel (B) of Figure~\ref{fig:steps}.
It is important to recall at this point that the number of events observed in each of the data bins follows a Poisson process.
Thus, it is not possible to project the data into the energy description, without destroying the Poisson nature of the histogram.
Instead it is necessary to convert the model into the data space (ADC channels).
To achieve this, the edges of the measured histogram, in ADC channel values ranging from 0 to 1024, were converted to energy by applying the calibration function in Equation~\ref{eq:calibration}.
The resulting width of the energy bins will not be constant.

The model templates are re-binned from their uniform energy grid to the detector's non-linear, energy-converted, ADC channel grid using the re-binning methods implemented in the becquerel python package \cite{becquerel}.
Panel (C) in Figure~\ref{fig:steps} shows the result of this operation using the set of parameters that best aligns the templates to the data.
This procedure aligns the bin edges of the simulated components with the measured data and properly tallies the amount of radiation expected in each histogram channel.

The subsequent step is to compute the 6 coefficients (weights) that best match the total weighted contribution of the templates to the observed data.
To retain the Poisson nature of the solution, we employ an Iteratively Reweighted Least Squares (IRLS) algorithm, which is known to converge to the Poisson maximum-likelihood solution \cite{Green1984}.
At each iteration, the model components and the data are re-weighted by the inverse of the uncertainty, which is defined as $\sigma_i = \mathrm{maximum}(1, \sqrt{\lambda_i})$, where $\lambda_i$ is the model prediction for bin $i$ from the previous iteration.
The non-negative least-squares solver is called on the weighted arrays.
After 4 iterations the solution was observed to reach a precision of $\approx 10^{-8}$.

The low-energy region of the spectrum (below a few hundred keV) is heavily influenced by scattering and absorption processes.
Consequently, the simple model used for creating the spectral components, which does not incorporate the surrounding geometry and only partially models the details of the detector housing, fails to accurately reproduce the measured spectral shape in this range.
The templates also do not model the high energy spectrum (above 3\,MeV) well, where the shape is entirely dominated by cosmic radiation.
To address these issues, the range where the data are compared to the model is limited to the range between a lower threshold $E_{\mathrm{min}}$ and upper threshold $E_{\mathrm{max}}$.
This energy range does not align perfectly with the ADC bin edges.
Our algorithm therefore identifies the specific data bins that contain $E_{\mathrm{min}}$ and $E_{\mathrm{max}}$.
It then calculates the fractional contributions of these boundary bins, effectively splitting them to match the energy range precisely.
Data counts outside this range are tallied into overflow terms, which are used for the regularization term that will be described in the next Section.
The result of the IRLS procedure, limited to that energy range, is shown in panel (D) of Figure~\ref{fig:steps}.

At this point the model describes the data as well as possible for a given energy resolution, cosmic power law, and a set of calibration parameters.
It remains to find the optimal values for these parameters.

\subsection{Parameter optimization}
\label{sec:param_opt}

There are 5 parameters that need to be found simultaneously; the multiplicative factor in Equation~\ref{eq:resolution} $\alpha$, the power law factor for modeling the cosmic component $\gamma$, the gain $g$, saturation $k$ and offset $o$, describing the calibration function.
The data matching procedure, outlined in the previous section, is assumed as a black-box function, returning an optimized model prediction for each of the bins in the data spectrum, given a set of these 5 parameters.
As mentioned previously, the observed data, in the form of a histogram, are governed by Poisson statistics.
As such, a natural loss function for the parameter optimization problem is the deviance~\cite{mccullagh_generalized_1989}, a negative log likelihood-derived statistic defined as:
\begin{equation}
\label{eq:deviance}
    D(\mathbf{n}|{\boldsymbol\lambda}) = 2 \sum_{i} n_i\log n_i - n_i \log \lambda_i + \lambda_i - n_i
\end{equation}

The variable $n_i$ refers to the $i$-th bin content in the measured spectrum and the variable $\lambda_i$ to the $i$-th bin model prediction, after the data matching procedure in the previous section was performed.
As a standard statistical metric, the deviance behavior is well understood, allowing for reliable interpretation of the result;
According to Wilks' theorem, this distribution should approximate a $\chi^2$ distribution centered around the number of degrees of freedom.
This number is typically calculated as the number of free parameters (here the number of bins) minus the number of parameters in the model (here 6 weights and 5 calibration parameters).
If $D(\mathbf{n}|{\boldsymbol\lambda})$ is normalized by that number of degrees of freedom, the resulting distribution will be centered at 1.
Wilks' theorem is not valid if the spectrum has bins with only a small number of events and the distribution will deviate from $\chi^2$ and typically be centered at a value slightly above 1.
The index $i$ in Equation~\ref{eq:deviance} will only be evaluated for bins between the lower and upper energy thresholds at ($E_{\mathrm{min}}$, $E_{\mathrm{max}}$).

The calibration problem is therefore framed as a 5-parameter optimization task, governed by a statistical meaningful loss function.
Various global and local optimization algorithms implemented in the non-linear solver package nlopt \cite{nlopt} were tested for their ability to find the correct solution.
It was found that a 3-stage initialization procedure, combining global and local searches, leads to stable convergence.

\begin{enumerate}
    \item Global search: A Controlled Random Search (CRS) algorithm \cite{CRS2} optimizes the 3 primary parameters: energy resolution ($\alpha$), cosmic index ($\gamma$), and gain ($g$). To increase speed, this stage uses a reduced set of background templates, excluding the 511\,keV and radon components.
    \item Local search, no offset: The result from the first stage is used as the initial guess for a local Subplex algorithm \cite{SUBPLEX}. This stage optimizes 4 parameters ($\alpha$, $\gamma$, $g$, $k$) using the full set of background templates, but with the offset $o$ fixed at 0.
    \item Local search, full model: The result from the second stage is used as the initial guess for a final Subplex search, which optimizes all 5 parameters ($\alpha$, $\gamma$, $g$, $k$, $o$) using the full model.
\end{enumerate}
This approach ensures a stable and reproducible convergence by first establishing the primary calibration parameters before refining the fit with the more degenerate saturation and offset terms. Some tweaks to the optimization routines were applied during each stage.

In the first stage, the deviance loss was altered and extended by a regularization term $O_\mathrm{R}$. The regularization term controls the amount of data that is outside the energy thresholds, by summing the bins that would fall above the upper energy thresholds and adding them them to the loss function. This prevents the model from simply matches a few bins in the highest energy region:

\begin{equation}
\label{eq:regularization}
(D(\mathbf{n}|{\boldsymbol\lambda}) + O_\mathrm{R} \log O_\mathrm{R}) / \sum_i n_i, \quad O_\mathrm{R} = \sum_{E(i)>E_{\mathrm{max}}} n_i
\end{equation}

The loss function is relatively flat with respect to the saturation and offset parameters.
Consequently, the global minimum is often governed by small variation in the data rather than the true underlying curvature.
However, the offset is expected to be close to zero.
The second stage ensures that the optimum is always approached along the same ``path'' in parameter space (with an offset close to zero), but it also means that the convergence toward a local minimum might consistently be on the same ``side'' of the true set of parameters, thus introducing a small bias (see Sec.~\ref{sec:results}).

The CRS and Subplex algorithms require lower and upper bounds and an initial guess, which were selected as outlined in Table~\ref{tab:bounds}.
The limits for the energy resolution ($\alpha$) were chosen to cover typical ranges for NaI(Tl) detectors.
The bounds for the cosmic power-law coefficient were set higher than the physically observed values ($\approx 1.3$ \cite{sandness_accurate_2009}). 
This adjustment accounts for the model's observed preference for high coefficients, a likely modeling artifact that serves to compensate for missing down-scattering in the templates.
The gain limits were based on the typical MeV range of gamma ray spectra and chosen to ensure at least 200 spectral bins are always present.
The saturation and offset limits are purely observational but provide sufficient flexibility for the model to adjust.
Finally, limits in later fitting stages were tightened to constrain the optimization, preventing the true parameters from exiting the bounds while reducing the convergence time.

\begin{table*}
    \centering
    \caption{Bounds and guesses for the default 3-stage optimization.
    ``$x_f$'' denotes the best-fit value from the previous stage for parameter $x$. 
    $E_\mathrm{max} = 2850$\,keV, $b_\mathrm{min}$ is the ADC channel of the 200th bin, and $b_\mathrm{max}$ is the highest ADC channel.
    }
    \label{tab:bounds}
    \begin{tabular}{c c c c}
        \hline
        Param. & Stage 1 (Global, CRS) & Stage 2 (Local, No Offset) & Stage 3 (Local, Full) \\
        \hline
        $\alpha$ (Res.) & [0.04, 0.25] & [$\alpha_f - 0.05, \alpha_f + 0.05$] & [$\alpha_f - 0.05, \alpha_f + 0.05$] \\
        $\gamma$ (Cosmic) & [1.5, 3.0] & [$\gamma_f - 0.5, \gamma_f + 0.5$] & [1.5, 3.0] \\
        $g$ (Gain) & [$b_\mathrm{min}/E_\mathrm{max}$, $b_\mathrm{max}/E_\mathrm{max}$] & [$0.7 \cdot g_f, 1.3 \cdot g_f$] & [$0.7 \cdot g_f, 1.3 \cdot g_f$] \\
        $k$ (Sat.) & Fixed (N/A) & [-1e-4, 1e-4] & [-1e-4, 1e-4] \\
        $o$ (Offset) & Fixed (N/A) & Fixed at 0 & [-20, 20] \\
        \hline
        Initial Guess & $\alpha=0.1, \gamma=2.5$ & $k=0.0$ & $o=0.0$ \\
        (for new params) & $g = \sqrt{b_\mathrm{min}b_\mathrm{max}}/E_\mathrm{max}$ & & \\
        \hline
    \end{tabular}
\end{table*}

\subsection{Continuous real-time calibration}
\label{sec:realtime_calib}

For real-time operation where spectra are processed sequentially, the full 3-stage initialization is not repeated.
Instead, a simplified and much faster procedure is used.
The procedure performs only a single-stage local optimization using the Subplex algorithm, identical to the third stage of the full initialization, optimizing all 5 parameters.
The method leverages the fact that calibration parameters typically drift slowly, so the final parameters from the previously calibrated spectrum are used as the initial guess for the current spectrum.
To further reduce the convergence time, the 6 component weights from the previous fit are passed as an initial guess to the IRLS solver.
The routine implements an adaptive bound-tracking logic.
After each successful fit, the bounds for the gain ($g$) and energy resolution ($\alpha$) parameters are re-centered around their new best-fit values ($g_f$ and $\alpha_f$), as shown in Stage 3 of Table~\ref{tab:bounds}.
The bounds for the other parameters (e.g., $\gamma, k, o$) remain fixed.
This dynamic bounding allows the calibration to ``track'' slow drifts in the detector's response over time.
As such, a spectrum can be calibrated in about ten seconds, using the CPU only of an edge deployable computer, such as the Jetson Xavier NX series \cite{nvidia_jetson_xavier}.

Another real-time complication is the potential presence of an unaccounted-for radioactive source, which can drastically alter the spectral shape. 
While many nuisance sources produce signals below the energy threshold, thus not impacting the calibration procedure, some sources leave detectable signals within the calibration range.
In such cases, the existing templates are insufficient to capture the additional radiation, leading to a deterioration of the loss function.
If the loss function exceeds a defined limit ($>$1.5), the calibration procedure is immediately halted.
A calibration function from a previous, stable iteration is then used until the loss function returns to its expected range.

\section{Results}
\label{sec:results}

The method outlined in the previous section was evaluated using three distinct approaches, which are discussed in the following sections.
First, the calibration routine was tested on simulated spectra, which allowed for a direct comparison against the known ground truth values.
Next, results from measurements taken within an environmental chamber are presented. These tests utilize real data but cover a temperature range wider than that which is typically encountered upon deployment in an urban setting.
Finally, the method was applied to data gathered in the field is shown, demonstrating its performance and robustness in a realistic urban deployment scenario.

\subsection{Simulations}
\label{sec:simulations}

\begin{figure}
  \centering\includegraphics[width=1.0\myfigurewidth]{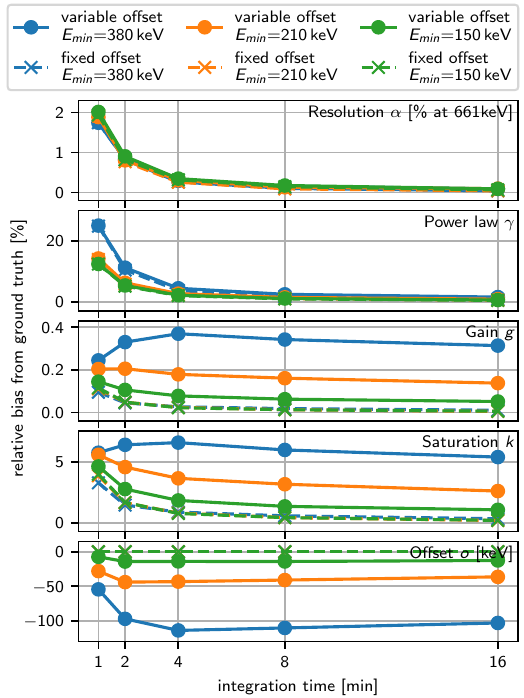}
  \caption{The observed mean bias of the optimization parameters for different integration times, $E_{min}$ and variable vs fixed offset. It is expressed as a percentage deviation from the ground truth average (7.5\% for the energy resolution, 2.2 for the cosmic power law, 0.275 for the gain, $2\times10^{-5}$ for the saturation and 2\,keV for the offset). The data for fixed offset lie all on top of each other and thus are barely distinguishable.}
  \label{fig:sim_bias}
\end{figure}

\begin{figure}
  \centering\includegraphics[width=1.0\myfigurewidth]{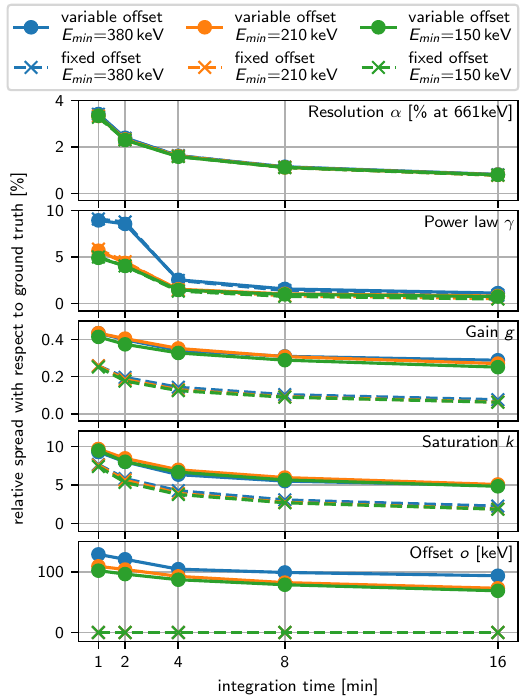}
  \caption{Same as Figure~\ref{fig:sim_bias}, but for the normalized standard deviations describing the spreads of the fitted parameters. It is expressed as a percentage spread normalized by the ground truth average. The data for fixed offset lie all on top of each other and thus are barely distinguishable.}
  \label{fig:sim_spread}
\end{figure}

Simulations were performed for 15 scenarios spanning five dwell times (1, 2, 4, 8, 16\,minutes) and three low energy threshold values (150\,keV, 210\,keV and 380\,keV). 
The upper energy threshold was fixed at 2850\,keV.
Each scenario consists of 20,000 spectra, that were based on the calibration model using well defined parameters and weights.
The six spectral components are weighted such that the resulting shape closely resembles data measured with a 2$\times$4$\times$16\,inch NaI(Tl) detector: 25,000 (cosmics), 1,000 (511\,keV), 15,000 (K-40),  1,500 (Rn-222 progeny), 12,500 (Th-232) and 16,000 (U-238) counts per minute (CPM).
The optimization parameters were uniformly sampled within realistic detector ranges: the resolution ($\alpha$) between 5\% and 10\%, the cosmic power law coefficient ($\gamma$) between 2.0 and 2.4, the gain ($g$) between 0.2 and 0.35, the saturation ($k$) between 0 and $4\times10^{-5}$ and the offset ($o$) between -4 and 4\,keV.

The calibration parameters were reconstructed for each spectrum using the procedure outlined in Section~\ref{sec:methods}.
The difference between the reconstructed parameters and the ground truth values was calculated for each realization.
The mean difference is considered the bias of the model, the standard deviation of the difference distribution is the spread.
The reconstruction performance is displayed in Figures~\ref{fig:sim_bias} and~\ref{fig:sim_spread}.
The most significant error is observed for the offset value ($o$), which exhibits a systematic bias of up to 2\,keV (solid blue line).
This bias is strongest when the fit is limited to higher energies; providing more low-energy bins constrains the offset and reduces the bias. Unfortunately, for measured data the low-energy region is not well modeled by the templates, restricting $E_{min}$ to the 200--300\,keV range.
Increasing the dwell time reduces this bias as well. 
Notably, the bias is least significant when the exact value of the offset is provided to the optimization routine as a known parameter.
Given that we sampled between -4 and 4\,keV, the spread of the offset of about 2\,keV, displayed in Figure~\ref{fig:sim_spread}, suggests that the offset is effectively unconstrained by the routine and the actual value found is mostly statistical fluctuation over the sampling range.
In contrast, the gain ($g$) and saturation ($k$) parameters are reconstructed with high fidelity, with gain bias usually less than 0.5\% and saturation less than 5\%.
Figure~\ref{fig:sim_spread} shows that while the offset distribution is relatively broad, the other parameters are recovered with high precision.
The saturation has a percentage standard deviation from the ground truth between 5--10\%, largely independent of whether the offset is constrained or not.
The gain has a percentage standard deviation of less than 0.5\%, indicating the correct value is reconstructed robustly.
The precision of the resolution ($\alpha$) and power law ($\gamma$) parameters are not the primary focus for calibration procedures, but their values are close to ground truth with long dwell times. The distributions also narrow with longer dwell times, showing that even these two ``secondary'' parameters are sufficiently constrained and closely reconstructed.

\begin{figure}
  \centering\includegraphics[width=1.0\myfigurewidth]{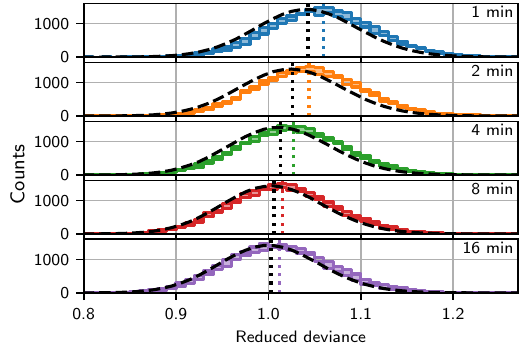}
  \caption{The distribution of the loss function (reduced deviance) for various dwell times: The colored shapes are from simulated spectra (the width of the bands indicates differences between various low-energy thresholds), the dashed lines are from the fixed-parameter fit (the statistical expected distribution under perfect circumstances). The vertical dotted lines mark the mean of each distribution.}
  \label{fig:sim_chi2}
\end{figure}

To investigate the underlying cause of the offset bias, a control experiment was performed using a ``fixed-parameter'' fit.
For all 20,000 simulated spectra, the five calibration parameters were fixed to their known ground truth values, and only the six component weights were fit.
This allows us to separate the impact of the template fitting procedure from the performance of the optimization routine and compare the results to the absolute statistical limit.
As discussed in Section~\ref{sec:param_opt}, the loss function, expressed by the reduced deviance, is expected to follow a $\chi^2$ distribution centered at one.
This expectation breaks down when bins with few counts are present and Wilks' theorem is not strictly valid for Poisson distributed data. Longer dwell times should mitigate this effect and yield deviance distributions that better match the statistical prediction.

Figure~\ref{fig:sim_chi2} shows the distribution of the observed loss function for different dwell times.
The small spread within the colored bands indicates that the low-energy threshold does not meaningfully affect the quality of the fit, meaning the model remains statistically consistent across different energy ranges.
Longer dwell times (bottom) yield distributions that better match the expected $\chi^2$ distributions, confirming that the trend is statistical in nature.

However, a small bias exists between the fixed-parameter fit (dashed lines) and the reduced deviance resulting from the full optimization procedure (colored shapes).
This indicates that the optimization procedure not always finds a unique solution matching the ground truth.
Instead, ambiguities between the gain ($g$), saturation ($k$), and offset ($o$) parameters occasionally cause the routine to settle in a local minimum where the offset is closer to zero than the actual ground truth.
This behavior is expected for a complex, multi-step optimization with a noisy criterion near the optimum. Because the procedure searches for parameters sequentially, it systematically prefers solutions closer to the initial guesses (which are zero for both saturation and offset).
While this situation is suboptimal from a purely statistical perspective, it acts as a form of implicit regularization towards the initial guess, ensuring the offset does not deviate far from zero in the absence of strong low-energy constraints.

\subsection{Measurements in an environmental chamber}
\label{sec:env_chamber}

\begin{figure}
  \centering\includegraphics[width=1.0\myfigurewidth]{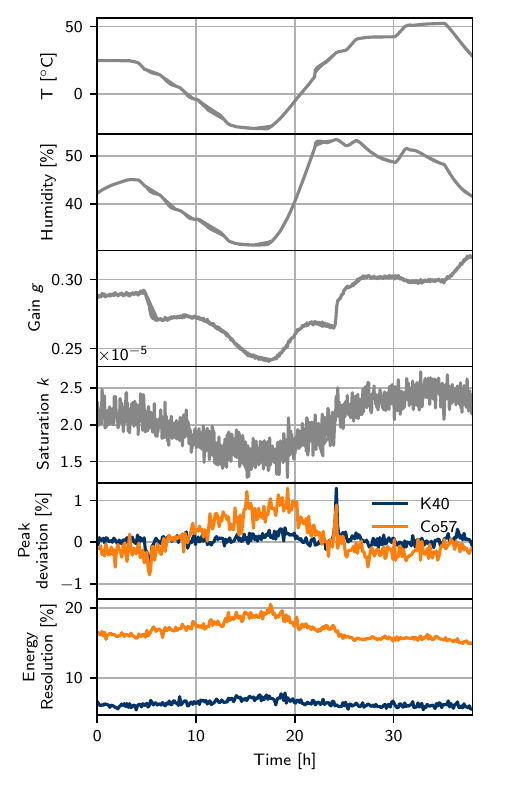}
  \caption{Results from temperature testing in an environmental chamber at ANL. The panels show, 
from top to bottom:
(1) The enclosure temperature vs. time (-25$^\circ$C to 50$^\circ$C).
(2) The enclosure relative humidity, which is mostly correlated with temperature.
(3) The fitted gain $g$, showing a non-linear correlation with temperature.
(4) The fitted saturation $k$.
(5) The peak position deviation (\%) for the 1460\,keV K-40 (in-range) and 122\,keV Co-57 lines.
(6) The fitted energy resolution (\%) for both lines, showing degradation at
low temperatures.}
  \label{fig:temp}
\end{figure}

As previously highlighted, the gain is particularly susceptible to changes in ambient temperature.
In an urban environment, detectors need to be operated over a large range of ambient temperatures.
To verify that the developed calibration algorithm is robust over a wide range of temperatures, the system was operated in an environmental stress test chamber for military grade certification (MIL STD 810G \cite{ELITE2026}), with a $-40^\circ$C to $+50^\circ$C temperature sweep, under various relative humidity conditions creating damp to freezing environments.
For this test, the continuous real-time calibration procedure (described in Section~\ref{sec:realtime_calib}) was used, operating on a 4-minute rolling buffer, performing the calibration procedure once every minute.
The fit was restricted to the energy region between 380\,keV and 2800\,keV.
This range is slightly different from the ranges explored in the previous section.
The DAQ offset is expected to be stable (not vary in time) and results in Sec.~\ref{sec:simulations} found it to be largely unconstrained by the optimization procedure.
As such the offset $o$ was fixed at 2.4\,keV for this analysis.
This value was established as a realistic DAQ offset by analyzing the 59.5\,keV line of Am-241 and the 122\,keV line of Co-57 and extrapolating to zero.

A Co-57 source was placed at a fixed standoff from the detector enclosure to provide a reference low-energy gamma line at 122\,keV.
Both the 1460\,keV K-40 and the 122\,keV Co-57 peaks are included in the presented analysis.
The first peak is used to check the calibration quality within the considered energy range, the second one is used to check the stability of the calibration at low energy outside the fitting range.

Figure~\ref{fig:temp} plots the fitted parameters as a function of time during the temperature testing. The top panel shows the temperature evolution as it ramps down to -25$^\circ$C, up to +50$^\circ$C, and back to ambient temperature.
The relative humidity (second panel) correlates with temperature but stays between 30-60\%.
The gain, $g$, displayed in the third panel, closely tracks the temperature curve.
It exhibits two abrupt changes at the 6-hour and 24-hour marks.
These rapid changes persists even when each spectrum is calibrated independently without using the previous state as an initial guess.
This consistency, combined with the fact that transitions occur at nearly identical temperatures, suggests a physical origin --- likely crystal relaxation --- rather than an artifact of the calibration routine.
The gain, $g$ (third panel), and saturation, $k$ (fourth panel), both vary smoothly with temperature.
However, the gain exhibits a non-linear correlation, which is expected for NaI(Tl) and suggests that a simple linear temperature correction would be insufficient.
The saturation parameter changes more modestly and linearly with temperature and the spread (noise) is considerable, at the order of 10\%.
Overall, a slight lag is visible between the temperature and the calibration parameters, which is expected from the thermal inertia of the crystal.

To quantify the stability of the final calibration, the centroid and FWHM values were extracted for the 1460\,keV (K-40) and 122\,keV (Co-57) lines using 8-minute data intervals.
These parameters were derived by fitting a Gaussian peak atop a linear background model, with the FWHM calculated directly from the resulting peak width.
The fifth panel of Figure~\ref{fig:temp} shows the peak position deviation for both lines.
The deviation remains less than 1\% for both peaks over the entire temperature sweep (e.g., a shift of less than 1.2\,keV for the 122\,keV line).
The largest excursion occurs when the temperature is below -20$^\circ$C and at 24 hours into the measurement where the best-fit gain parameter appears to temporarily converge to a false minimum.   
The stability observed at low energy shows that the estimated calibration parameters apply reasonably well in that region as well, even though outside the fit range.
The sixth panel of Figure~\ref{fig:temp} shows the fitted energy resolution for both peaks. The resolution clearly degrades as the temperature drops.
This effect reverses as the detector warms back up, returning to its baseline value at and above ambient temperature, where it remains stable.

\begin{figure}
  \centering\includegraphics[width=1.0\myfigurewidth]{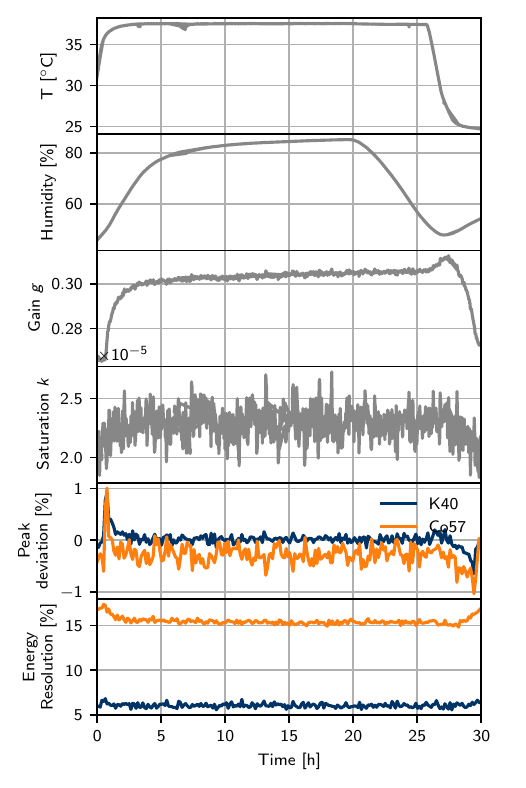}
  \caption{Same as Figure~\ref{fig:temp}, but for varying relative humidity.}
  \label{fig:hum}
\end{figure}

Figure~\ref{fig:hum} repeats the panels shown in Figure~\ref{fig:temp}, but with the environmental chamber set to hold a temperature of 35$^\circ$C while the relative humidity is slowly ramped up and maintained at 85\%, before a slow ramp down. The temperature changes from ambient to 35$^\circ$C and back to ambient occur relatively fast compared to the relative humidity changes.
Once again, the gain trends to lag the temperature behavior. The peak deviation and fitted energy resolution are both largest during the fast transitions in temperature.
The humidity test, however, shows that the relative humidity has no appreciable impact on the detector, with the peak deviation, as well as the resolution remaining flat for the remaining parts of the test.
Even during the temperature transient, the peak deviation remains below 1\%.

\subsection{Performance in the field}
\label{sec:field_performance}

\begin{figure}
  \centering\includegraphics[width=1.0\myfigurewidth]{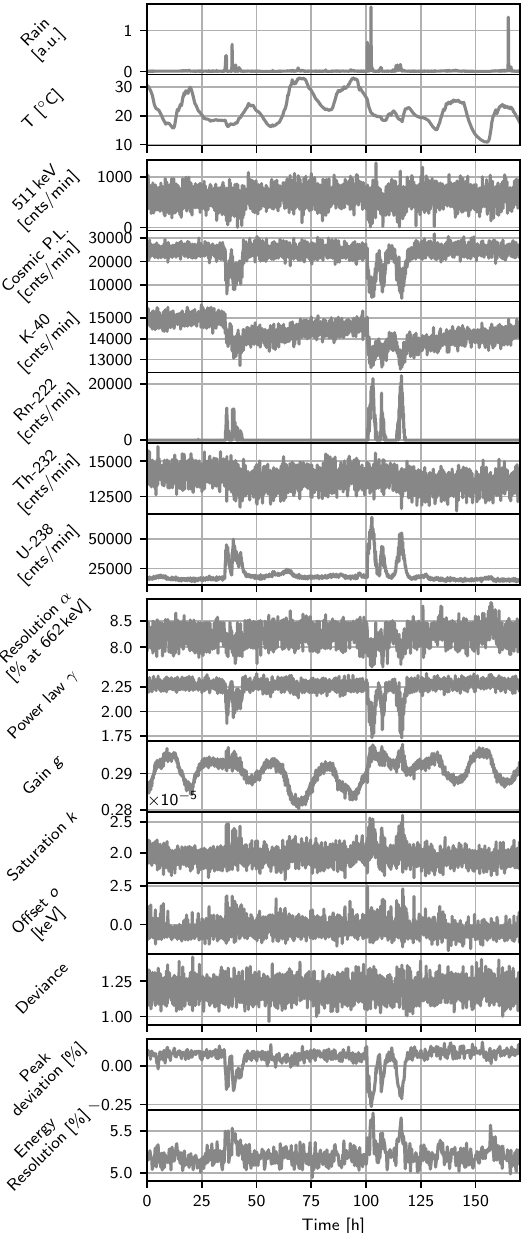}
  \caption{Time series of all key parameters of the week long field trial: The top block shows rain fall and temperature. The second block shows the fitted weights of the spectrum components. Significant increases in the Rn-222 progeny and U-238 components are clearly correlated with periods of rain. The third block shows the key calibration parameters and the reduced deviance. Rain does not significantly affect these parameters and the deviance is stable for the entire period. The last block shows stability metrics for the primary K-40 1460\,keV peak. A stable peak deviation ($\pm 0.25\%$ range) and energy resolution confirms that the continuous calibration procedure maintains the energy scale despite environmental changes.}
  \label{fig:chicago_all}
\end{figure}

To test the stability and robustness of the presented calibration procedure under realistic, continuously changing environmental conditions, one of the PANDA systems was deployed at ANL in 2022, for approximately 7\,days of operation.
To ensure the stability of the initialization procedure, the latter was applied to all  2,553 consecutive 4-minute integrated spectra in the dataset.
The calibration routine was executed with a fixed fit range between 380 and 2,850\,keV.
During the procedure, no prior assumption was made on the range of possible calibration parameters, except those fundamental constraints outlined in Table~\ref{tab:bounds}.

One particular characteristic of naturally occurring radioactive materials present in urban environments is a strong increase in the activity of the Rn-222 progeny during rain events due to washout from the atmosphere.
The calibration routine is designed to be stable in these conditions, using a special Rn-222 progeny component to account for the expected changes during rain. The second block in Figure~\ref{fig:chicago_all} presents the time series of the fitted weights for all six components throughout the week-long calibration test.
During rain events (e.g., about 40 hours and 100 hours in), a pronounced spike is observed in the fitted weights for both the Rn-222 progeny and the U-238 components.
Rn-222 is part of the U-238 decay chain but is also a noble gas, which allows it to migrate into the atmosphere and be separated from its parents.
The simulated components for the Rn-222 progeny and U-238 are similar in shape, which can cause the fit amplitude of the U-238 component to rise alongside the Rn-222 progeny amplitude during rain events.
The cosmic power law parameter $\gamma$ is shown in the 2nd panel of the third block in Figure~\ref{fig:chicago_all} and also varies during rain events.
This indicates that while the Rn-222 progeny template was carefully crafted, the actual spectral changes during a rain event are not entirely absorbed by that template and require modifications to other components' weights and shapes in ways that are unlikely to be physical.
However, the model is sufficiently flexible to compensate for these changes and still achieve a good overall fit, as reflected in the deviance metric, which is largely constant and centered at approximately 1.2.

The third block in Figure~\ref{fig:chicago_all} shows the time evolution of the key calibration parameters: gain, saturation, and offset.
As expected, the gain is anti-correlated with daily temperature fluctuations, with lows and highs of about $10^{\circ}$C and $35^{\circ}$C during the trial.
Specifically, the gain remains within a tight range of approximately 0.28 to 0.29.
The saturation and offset are mostly constant, showing little influence from the diurnal temperature variations.
In particular, the saturation remains centered near 2.0, with its temporal profile dominated more by statistical spread (noise) rather than by systematic variations over time.
The offset is similarly stable, staying centered near 0.0 keV.

To verify the stability of the energy scale, the K-40 line at 1460\,keV is fit with a Gaussian peak atop a linear background and displayed in the last block of Figure~\ref{fig:chicago_all}.
The peak deviation remains constrained within the $\pm 0.25\%$ range, and the FWHM energy resolution at 1,460\,keV hovers around 5.2\%.
Although modest increases in the K-40 resolution are visible during rain, the resolution parameter of the calibration routine, $\alpha$, actually decreases slightly during these periods.
This discrepancy suggests that the apparent increase in the K-40 peak's width most likely originates from a changing spectral background shape that affects the stability of the simple Gaussian-plus-linear fit model, rather than a true degradation in instrumental resolution or calibration stability.
The small extent of these fluctuations provides strong evidence that the continuous calibration effectively maintains an accurate energy scale despite significant environmental stressors over the week long period.
Furthermore, the successful processing of all 2,553 spectra demonstrates that the calibration procedure is robust to a realistic range of environmental and instrumental variations.

\section{Discussion}
\label{sec:discussion}

The calibration method presented in this work 
relies on a physically-motivated, multi-parameter model rather than a simple empirical function. It provides a robust, automatic, and continuous energy calibration for NaI(Tl) detectors operating in unattended, real-world scenarios.
In particular, results from the field deployment, described in Section~\ref{sec:field_performance}, show that the method effectively decouples changes in environmental gamma-ray flux, i.e., during rain events, from the temperature-induced instrumental drift.
During this deployment, the gain parameter (see Figure~\ref{fig:chicago_all}) tracked the ambient temperature without being affected by the rain events, and the K-40 1460 keV peak position remained stable to within $\pm 0.25\%$.
Furthermore, testing in an environmental chamber (see Section~\ref{sec:env_chamber}), shows that the model, fitted above 380\,keV, correctly extrapolates at low energies far below the fitting range, maintaining the Co-57 122\,keV peak to within 1\% over a $75^\circ$C temperature swing (see Figure~\ref{fig:temp}).
While the offset was fixed in this test, aiding the model to remain stable at low energies, the field deployment did not make that assumption, and still resulted in a tightly constrained offset in the $\pm\,1$\,keV range.
Furthermore, the full initialization routine, with limited constraints on the parameters, reliably found a reasonable solution for all 2553 spectra in that test, further illustrating the robustness of the method.
All these results provide confidence that the calibration is accurate across the entire energy spectrum, from the low-energy region relevant for many illicit sources to the high-energy NORM lines, and that the method is sufficiently flexible to work in diverse environments.

A significant observation arises when comparing the fitted gain parameter $g$ between the field and environmental chamber tests.
The two systems, despite being of the same model, exhibit different, opposing responses to temperature.
In the field deployment (see Figure~\ref{fig:chicago_all}), the gain is anti-correlated with temperature.
This observation is consistent with thermal quenching in scintillators, where the light output of NaI(Tl) decreases as temperature rises \cite{ianakiev2006temp}.
The environmental chamber test (Figure~\ref{fig:temp}), however, reveals a positive correlation.
The gain $g$ increases with temperature across the entire $75^\circ$C range.
This positive correlation is contrary to expectations, but as observed by Ianakiev et al.\ in \cite{ianakiev2006temp}, the temperature not only affects the total light output, but also the fraction of photons in the fast and slow components of the light curve.
With more photons being shifted to the slow component at decreasing temperature, they will not be captured by the shaper, unless the shaping time is set sufficiently long.
In this case, fewer photons contribute to the observed signal, resulting in a reversal of the effect and a correlation between temperature and gain, especially far below ambient temperature.
Our shaping time was set at 0.9\,$\upmu$s, which is similar to the range where Ianakiev et al.\ observed this behavior.
Temperature also affects the performance of PMTs as described early on by Young et al.\ in \cite{young1963temp}, with typically higher amplification factors at higher temperatures.
This suggests that competing effects may be at play.
Furthermore, the relatively compact PMT cools down more quickly compared the NaI(Tl) crystal with its large thermal inertia.
This finding highlights the presence of component-level variation that can exist even in integrated detector systems from the same manufacturer, and stresses the necessity of our approach.
A simple, universal temperature correction formula cannot be applied to these detectors; the model must be flexible enough to track the net gain $g$ as an independent, time-varying parameter.

Our approach offers significant advantages over conventional methods.
It works fully at a software level, so it can be applied to any system with a digital readout.
It does not require any hardware; particularly, it obviates the need for complex and energy-intensive heating and/or cooling systems for temperature control.
When compared to simple peak-locking algorithms, the method is more robust as it uses information from the entire spectrum rather than relying on one or two prominent, but potentially fluctuating, NORM peaks.
It also resolves the issue of manual configuration as the initialization procedure only imposes limited constraints, which do not require configuration from an operator.

There are some limitations and areas for improvement.
The simulation study, presented in Section~\ref{sec:simulations}, revealed that the three-stage optimization procedure introduced to resolve a degeneracy in parameter space leading to very similar loss values, itself introduced a bias in the deviance, by consistently moving through the parameter space along a specific path.
The effect of the bias is particularly prominent for the saturation and offset parameters, which are not well constrained within the fitting range.
However, this systematic bias does not appear to be a significant flaw of the approach.
The field test in Section~\ref{sec:field_performance} did not constrain the offset parameter, but still showed stable performance, indicating that the procedure leads to a solution sufficiently close to reality.
A second limitation was identified in the field data during rain events.
The observation that the U-238 and cosmic component weights, and the cosmic index $\gamma$, change along the Rn-222 progeny component weight (see Figure~\ref{fig:chicago_all}) implies that the radon template cannot capture the full spectral change during rain events, and that there is likely some crosstalk with other components of the model.
In particular, the fitted power law values are quite high, compared to more realistic values from other publications (such as 1.3 in \cite{sandness_accurate_2009}).
However, this result simultaneously demonstrates the resilience of the method.
The model was flexible enough to absorb spectral discrepancies by adjusting other components.
This ability is demonstrated in Figure~\ref{fig:chicago_all}, where the fitted resolution of the K-40 peak shows a slight correlation with the rain, but the peak location remains stable within $\pm 0.25\%$, proving the calibration itself was not appreciably compromised.

The fitted background weights themselves, as shown in Figure~\ref{fig:chicago_all}, represent a valuable scientific by-product, providing a real-time, quantitative measurement of the NORM environment of the detector. Such information could be used for correlative environmental studies, such as mapping radon progeny washout, or even supporting anomaly detection and isotope identification by the system.

Future work could focus on refining the background simulations, particularly for radon, to improve the physical accuracy of the fitted weights. Additionally, investigating correlations between fit parameters may reveal non-physical dependencies. Resolving these dependencies through regularization or by removing parameter ambiguities will simplify the calibration routine and increase its robustness.
Investigating the use of the method for mobile systems, for which the variations in terrestrial components are much more pronounced, would be another interesting direction for research.
Some initial tests show promising results, suggesting that the method is reliable in such environments as well.
Finally, the method presented here was tested with a single type of detector.
However, with some tweaking of the underlying physics assumptions, it should be flexible enough to work on any type of detector.
Applying the method to other types of detector such as Cadmium Zinc Telluride (CZT) and High-Purity Germanium (HPGe) would be an interesting pursuit.

\section{Conclusion}
\label{sec:conclusion}

This work presents a novel software-based method for the continuous, automated energy calibration of NaI(Tl) detectors operating in dynamic environmental conditions.
By combining a full-spectrum analysis using background radiation templates with a detailed physical model of the detector response, the need for active temperature stabilization is eliminated.
The method was successfully validated through simulations, environmental chamber testing across a wide temperature range ($-25^\circ$C to $+50^\circ$C), and a multi-day field deployment.
Results demonstrated the ability to maintain calibration stability over significant temperature fluctuations and precipitation events, effectively decoupling instrumental drift from changes in the environmental radiation flux.
Crucially, this approach reduces the operational burden of manual calibration, and eliminates the need for complex temperature stabilization systems and characterization measurements.
Ultimately, by replacing simple empirical gain adjustments with this comprehensive software-based model, the method improves accuracy while minimizing deployment complexity.
Overcoming these usual operational constraints is essential to enable the practical, unattended operation of dozens or hundreds of distributed nodes across an urban landscape.

\bibliographystyle{IEEEtran}
\bibliography{IEEEabrv, ref}

@IEEEtranBSTCTL{IEEEexample:BSTcontrol,
  CTLuse_forced_etal = "yes", 
  CTLmax_names_forced_etal = "1",
}

@Inbook{Kouzes2020,
author="Kouzes, Richard",
editor={Fleck, Ivor
and Titov, Maxim
and Grupen, Claus
and Buvat, Ir{\`e}ne},
title="Radiation Detection Technology for Homeland Security",
bookTitle="Handbook of Particle Detection and Imaging",
year="2020",
publisher="Springer International Publishing",
address="Cham",
pages="1--31",
isbn="978-3-319-47999-6",
doi={10.1007/978-3-319-47999-6_50-1},
}

@book{
  international2005iaea,
  title={Environmental and Source Monitoring for Purposes of Radiation Protection},
  series={General Safety Guide},
  number={RS-G-1.8},
  year={2005},
  isbn={92-0-113404-5},
  url={https://www.iaea.org/publications/7176/environmental-and-source-monitoring-for-purposes-of-radiation-protection},
  author={{International Atomic Energy Agency}},
  publisher={{International Atomic Energy Agency}},
  address={Vienna}
}

@techreport{Mitchell2013GADRAS,
  author    = {Mitchell, Dean J. and Horne, Steven M. and Theisen, Lisa A. and Thoreson, Gregory G. and Harding, Lee T. and Bradley, Jon D. and Eldridge, Bryce D. and Amai, Wendy},
  title     = {{GADRAS-DRF User's Manual}},
  institution = {Sandia National Laboratories},
  year      = {2013},
  month     = {September},
  number    = {SAND2013-7503},
  doi       = {10.2172/1096506}
}

@ARTICLE{10247022,
  author={Bandstra, M. S. and Abgrall, N. and Cooper, R. J. and Hellfeld, D. and Joshi, T. H. Y. and Negut, V. and Quiter, B. J. and Salathe, M. and Sankaran, R. and Kim, Y. and Shahkarami, S.},
  journal={IEEE Transactions on Nuclear Science}, 
  title={Background and Anomaly Learning Methods for Static Gamma-Ray Detectors}, 
  year={2023},
  volume={70},
  number={10},
  pages={2352-2363},
  doi={10.1109/TNS.2023.3313996}}

@manual{nai_manual,
  title        = "Drawings for model 2X4H16/3A",
  organization = "Saint-Gobain Crystals Headquarters",
  address      = "17900 Great Lakes Parkway, Hiram, OHIO 44234-9681, USA",
  month        = "07",
  year         = "2022",
  note         = "Accessed on 11th July 2022",
  url          = "https://www.crystals.saint-gobain.com/sites/hps-mac3-cma-crystals/files/2021-09/s600-8391.pdf",
}

@manual{digitbase_manual,
  title        = "digiBASE 14-Pin PMT Base with Integrated Bias Supply, Preamplifier, and MCA with Digital Signal Processing ",
  organization = "ORTEC/AMETEK",
  address      = "801 South Illinois Avenue, Oak Ridge, Tennessee  37830, USA",
  month        = "07",
  year         = "2022",
  note         = "Accessed on October 2025",
  url          = "https://www.ortec-online.com/-/media/ametekortec/manuals/d/digibase-e-mnl.pdf",
}

@article{engelkemeir1956,
    author = {Engelkemeir, D. },
    title = {Nonlinear Response of NaI(Tl) to Photons},
    journal = {Review of Scientific Instruments},
    volume = {27},
    number = {8},
    pages = {589-591},
    year = {1956},
    doi = {10.1063/1.1715643},
    URL = {https://doi.org/10.1063/1.1715643},
    eprint = {https://doi.org/10.1063/1.1715643}
}

@article{Hull2009,
  author={Hull, G. and Choong, W. -S. and Moses, W. W. and Bizarri, G. and Valentine, J. D. and Payne, S. A. and Cherepy, N. J. and Reutter, B. W.},
  journal={IEEE Transactions on Nuclear Science},
  title={Measurements of NaI(Tl) Electron Response: Comparison of Different Samples},
  year={2009},
  volume={56},
  number={1},
  pages={331-335},
  doi={10.1109/TNS.2008.2009876}
}

@article{Valentine1998,
  author={Valentine, J. D. and Rooney, B. D. and Dorenbos, P.},
  journal={IEEE Transactions on Nuclear Science},
  title={More on the scintillation response of NaI(Tl)},
  year={1998},
  volume={45},
  number={3},
  pages={1750-1756},
  doi={10.1109/23.675868}
}

@article{Rooney1997,
  author={Rooney, B.D. and Valentine, J.D.},
  journal={IEEE Transactions on Nuclear Science},
  title={Calculating nonproportionality of scintillator photon response using measured electron response data},
  year={1997},
  volume={44},
  number={3},
  pages={509-516},
  doi={10.1109/23.603681}
}

@article{Dorenbos1995,
  author={Dorenbos, P. and de Haas, J.T.M. and van Eijk, C.W.E.},
  journal={IEEE Transactions on Nuclear Science},
  title={Non-proportionality in the scintillation response and the energy resolution obtainable with scintillation crystals},
  year={1995},
  volume={42},
  number={6},
  pages={2190-2202},
  doi={10.1109/23.489415}
}

@INPROCEEDINGS{sandness_accurate_2009,
  author={Sandness, Gerald A. and Schweppe, John E. and Hensley, Walter K. and Borgardt, James D. and Mitchell, Allison L.},
  booktitle={2009 IEEE Nuclear Science Symposium Conference Record (NSS/MIC)}, 
  title={Accurate modeling of the terrestrial gamma-ray background for homeland security applications}, 
  year={2009},
  volume={},
  number={},
  pages={126-133},
  doi={10.1109/NSSMIC.2009.5401843}}

@ARTICLE{bandstra_modeling_2020,
  author={Bandstra, M. S. and Joshi, T. H. Y. and Bilton, K. J. and Zoglauer, A. and Quiter, B. J.},
  journal={IEEE Transactions on Nuclear Science}, 
  title={Modeling Aerial Gamma-Ray Backgrounds Using Non-negative Matrix Factorization}, 
  year={2020},
  volume={67},
  number={5},
  pages={777-790},
  doi={10.1109/TNS.2020.2978798}}

@article{CASANOVAS201278,
title = {Energy and resolution calibration of NaI(Tl) and LaBr3(Ce) scintillators and validation of an EGS5 Monte Carlo user code for efficiency calculations},
journal = {Nuclear Instruments and Methods in Physics Research Section A: Accelerators, Spectrometers, Detectors and Associated Equipment},
volume = {675},
pages = {78-83},
year = {2012},
issn = {0168-9002},
doi = {https://doi.org/10.1016/j.nima.2012.02.006},
url = {https://www.sciencedirect.com/science/article/pii/S0168900212001490},
author = {R. Casanovas and J.J. Morant and M. Salvadó}
}

@misc{nlopt,
  title = {The {NLopt} nonlinear-optimization package},
  author = {Steven G. Johnson},
  year = {2007},
  howpublished = {\url{https://github.com/stevengj/nlopt}}
}

@article{CRS2,
  title = {Some variants of the controlled random search algorithm for global optimization},
  author = {P. Kaelo and M. M. Ali},
  doi = {10.1007/s10957-006-9101-0},
  year = {2006},
  volume = {130},
  pages = {253--264},
  journal = {Journal of Optimization Theory and Applications}
}

@phdthesis{SUBPLEX,
  author  = {Thomas Harvey Rowan},
  title   = {Functional stability analysis of numerical algorithms},
  school  = {Department of Computer Science, University of Texas at Austin},
  year    = {1990},
  address = {Austin, TX}
}

@misc{becquerel,
  author = {{Lawrence Berkeley National Laboratory}},
  title = {becquerel: A Python package for nuclear spectroscopy},
  howpublished = {\url{https://github.com/lbl-anp/becquerel}},
  year = {2025},
  note = {Accessed: October 2025}
}

@article{Green1984,
  author = {Green, P. J.},
  title = {Iteratively Reweighted Least Squares for Maximum Likelihood Estimation, and some Robust and Resistant Alternatives},
  journal = {Journal of the Royal Statistical Society. Series B (Methodological)},
  volume = {46},
  number = {2},
  pages = {149--192},
  year = {1984},
  doi = {10.1111/j.2517-6161.1984.tb01288.x},
  url = {https://rss.onlinelibrary.wiley.com/doi/abs/10.1111/j.2517-6161.1984.tb01288.x}
}

@misc{ianakiev2006temp,
    title={Temperature behavior of NaI (Tl) scintillation detectors},
    author={K. D. Ianakiev and B. S. Alexandrov and P. B. Littlewood},
    year={2006},
    eprint={physics/0605248},
    archivePrefix={arXiv},
    primaryClass={physics.ins-det}
}

@article{young1963temp,
    author = {Young, Andrew T.},
    journal = {Applied Optics},
    number = {1},
    pages = {51--60},
    publisher = {Optica Publishing Group},
    title = {Temperature effects in photomultipliers and astronomical photometry},
    volume = {2},
    month = {Jan},
    year = {1963},
    doi = {10.1364/AO.2.000051}
}

@article{Cooper2023,
doi = {10.1088/1742-6596/2586/1/012125},
url = {https://doi.org/10.1088/1742-6596/2586/1/012125},
year = {2023},
month = {sep},
publisher = {IOP Publishing},
volume = {2586},
number = {1},
pages = {012125},
author = {Cooper, R.J. and Abgrall, N. and Aversano, G. and Bandstra, M.S. and Hellfeld, D. and Joshi, T.H. and Negut, V. and Quiter, B.J. and Rofors, E. and Salathe, M. and Vetter, K. and Beckman, P. and Catlett, C. and Ferrier, N. and Kim, Y. and Sankaran, R. and Shahkarami, S. and Amitkumar, S. and Ayton, E. and Kim, J. and Volkova, S.},
title = {Networked Sensing for Radiation Detection, Localization, and Tracking},
journal = {Journal of Physics: Conference Series}
}

@article{Bandstra2026,
title = {Full spectrum modeling of in situ gamma-ray detector measurements with a focus on precipitation-induced transients},
journal = {Journal of Environmental Radioactivity},
volume = {291},
pages = {107826},
year = {2026},
issn = {0265-931X},
doi = {https://doi.org/10.1016/j.jenvrad.2025.107826},
url = {https://www.sciencedirect.com/science/article/pii/S0265931X25002139},
author = {M.S. Bandstra and J.M. Ghawaly and D.E. Peplow and D.E. Archer and B.J. Quiter and T.H.Y. Joshi and A.D. Nicholson and M.J. Willis and I. Garishvili and A.J. Rowe and B.R. Longmire and J.T. Nattress}
}

@article{Moszynski2002,
title = {Intrinsic energy resolution of NaI(Tl)11Support for this work was provided by the Polish Committee for Scientific Research, Grant No 8 T10C 002 20.},
journal = {Nuclear Instruments and Methods in Physics Research Section A: Accelerators, Spectrometers, Detectors and Associated Equipment},
volume = {484},
number = {1},
pages = {259-269},
year = {2002},
issn = {0168-9002},
doi = {https://doi.org/10.1016/S0168-9002(01)01964-7},
url = {https://www.sciencedirect.com/science/article/pii/S0168900201019647},
author = {M. Moszyński and J. Zalipska and M. Balcerzyk and M. Kapusta and W. Mengesha and J.D. Valentine}
}

@article{PFUND2016174,
title = {Improvements in the method of radiation anomaly detection by spectral comparison ratios},
journal = {Applied Radiation and Isotopes},
volume = {110},
pages = {174-182},
year = {2016},
issn = {0969-8043},
doi = {https://doi.org/10.1016/j.apradiso.2015.12.063},
url = {https://www.sciencedirect.com/science/article/pii/S0969804315304024},
author = {D.M. Pfund and K.K. Anderson and R.S. Detwiler and K.D. Jarman and B.S. McDonald and B.D. Milbrath and M.J. Myjak and N.C. Paradis and S.M. Robinson and M.L. Woodring}
}

@ARTICLE{8673769,
  author={Bilton, K. J. and Joshi, T. H. and Bandstra, M. S. and Curtis, J. C. and Quiter, B. J. and Cooper, R. J. and Vetter, K.},
  journal={IEEE Transactions on Nuclear Science}, 
  title={Non-negative Matrix Factorization of Gamma-Ray Spectra for Background Modeling, Detection, and Source Identification}, 
  year={2019},
  volume={66},
  number={5},
  pages={827-837},
  doi={10.1109/TNS.2019.2907267}
}

@misc{ELITE2026,
  author       = {{Elite Electronic Engineering, Inc.}},
  title        = {Environmental Stress Testing Chamber},
  year         = {2026},
  url          = {https://www.elitetest.com/testing-services/environmental-stress-testing/},
  urldate      = {2026-03-03},
}

@misc{nvidia_jetson_xavier,
  author       = {{NVIDIA Corporation}},
  title        = {Jetson Xavier Series},
  year         = {2026},
  url          = {https://www.nvidia.com/en-us/autonomous-machines/embedded-systems/jetson-xavier-series/},
  urldate      = {2026-03-03},
}

@book{mccullagh_generalized_1989,
  title         = {Generalized {Linear Models}, {Second Edition}},
  author        = {McCullagh, P. and Nelder, John A.},
  year          = 1989,
  publisher     = {{CRC Press}},
  isbn          = {978-0-412-31760-6},
  date          = {1989-08-01},
  langid        = {english},
  pagetotal     = 536,
}

@techreport{osti_1805000,
  author       = {Peplow, Douglas E. and Davidson, Gregory G. and Celik, Cihangir and Biondo, Elliott D. and Hackett, Alexandra C. and Ray, William R. and Archer, Daniel E. and Ghawaly, Jr., James M. and Nicholson, Andrew D. and Willis, Michael J. and others},
  title        = {Monte Carlo Simulation of Background and Source Measurements with CSG and CAD Geometries},
  institution  = {Oak Ridge National Laboratory (ORNL), Oak Ridge, TN (United States)},
  doi          = {10.2172/1805000},
  url          = {https://www.osti.gov/biblio/1805000},
  place        = {United States},
  year         = {2021},
  month        = {06}}

\end{document}